\documentclass[aps,prl,twocolumn,superscriptaddress]{revtex4-1}

\usepackage{subfigure}
\usepackage{graphicx}
\usepackage{bm}
\usepackage{amsmath}
\usepackage{epstopdf}
\usepackage{dcolumn}

\newcommand{\STO}{SrTiO$_3$}
\newcommand{\LAO}{LaAlO$_3$}
\newcommand{\Tc}{$T_c$}

\newcommand{\etal}{\emph{et al.}}

\begin{document}

\title{Vortex excitations in the Insulating State of an Oxide Interface.}
\author{M. Mograbi}
\affiliation{Raymond and Beverly Sackler School of Physics and Astronomy, Tel-Aviv University, Tel Aviv, 69978, Israel}
\author{E. Maniv}
\affiliation{Raymond and Beverly Sackler School of Physics and Astronomy, Tel-Aviv University, Tel Aviv, 69978, Israel}
\author{P. K. Rout}
\affiliation{Raymond and Beverly Sackler School of Physics and Astronomy, Tel-Aviv University, Tel Aviv, 69978, Israel}
\author{D. Graf}
\affiliation{National High Magnetic Field Laboratory, Tallahassee, Florida 32310, USA.}
\author{J. -H Park}
\affiliation{National High Magnetic Field Laboratory, Tallahassee, Florida 32310, USA.}
\author{Y. Dagan}
\affiliation{Raymond and Beverly Sackler School of Physics and Astronomy, Tel-Aviv University, Tel Aviv, 69978, Israel}
\email[]{yodagan@post.tau.ac.il}

\date{\today}

\begin{abstract}
In two-dimensional (2D) superconductors an insulating state can be induced either by applying a magnetic field, $H$, or by increasing disorder. Many scenarios have been put forth to explain the superconductor to insulator transition (SIT): dominating fermionic physics after the breaking of Cooper pairs, loss of phase coherence between superconducting islands embedded in a metallic or insulating matrix and localization of Cooper pairs with concomitant condensation of vortex-type excitations. The difficulty in characterizing the insulating state and its origin stems from the lack of a continuous mapping of the superconducting to insulating phase diagram in a single sample. Here we use the two-dimensional (2D) electron liquid formed at the interface between the two insulators (111) \STO~ and \LAO~ to study the superconductor to insulator transition. This crystalline interface surprisingly exhibits very strong features previously observed only in amorphous systems. By use of electrostatic gating and magnetic fields, the sample is tuned from the metallic region, where supeconductivity is fully manifested, deep into the insulating state. Through examination of the field dependence of the sheet resistance and comparison of the response to fields in different orientations we identify a new magnetic field scale, H$_{pairing}$, where superconducting fluctuations are muted. Our findings show that vortex fluctuations excitations and Cooper pair localization are responsible for the observed SIT and that these excitations surprisingly persist deep into the insulating state.
\end{abstract}

\maketitle

The superconductor to insulator transition is a prototypical quantum phase transition where the ground state of a 2D system transitions from a superconductor into an insulator upon changing a control parameter such as film thickness, disorder or magnetic field. This transition has been demonstrated in a variety of thin films such as bismuth \cite{HavilandandGoldmannSITthickness}, InO$_x$ \cite{HebardPaalanenInOx,ShaharInOxPeak,ovadia2013duality,Steiner2008Approach,KapitulnikInOxPeak}, MoGe \cite{KapitulnikMoGe}, TiN \cite{Baturina2007Quantum,BaturinaTiN}, cuprate superconductors \cite{Leng2011Electrostatic,BozovicLSCO} and more \cite{GoldmanVariedSIT,SITTantalum,SITNbSi,2eMROscillations,SITSnAndPb}.
\par
The many scenarios put forth to explain the SIT can be divided into two main categories. The fermionic scenarios suggest the insulating behaviour is the result of fermionic physics dominating after the breaking of Cooper pairs \cite{finkel1994superconducting,SITTheoryMirlin,MoC_STM}.
In contrast, in the bosonic scenarios the insualting state coincided with and is related to the existence of Cooper pairs. The two main theoretical ideas for the bosonic insulator are loss of phase coherence between superconducting islands embedded in an insulating matrix \cite{KapitulnikShimshoniQuantumMelts,dubi2007nature,Skvortsov2005Superconductivity} and localization of Cooper pairs with concomitant condensation of vortex excitations \cite{Fisher1990Presence,FisherSITTheory}.
\par
Many intriguing phenomena are observed in the SIT, such as scaling near a quantum critical point \cite{HebardPaalanenInOx,ovadia2013duality,Steiner2008Approach,KapitulnikMoGe,Leng2011Electrostatic,BozovicLSCO,SITTantalum}, large magnetoresistance peaks \cite{ShaharInOxPeak,Steiner2008Approach,KapitulnikInOxPeak,Baturina2007Quantum,BaturinaTiN} and thermally activated insulating behaviour \cite{ShaharInOxPeak,KapitulnikInOxPeak,BaturinaTiN,2eMROscillations}. However, some of these effects are not observed in all materials that exhibit a SIT, and a continuous tuning from the superconductor to the insulator state (where the sheet resistance becomes greater than $h/e^2$) in a single sample is still lacking. Both of these issues make the interpretation of the various phenomena controversial and there is no consensus regarding the mechanism of the SIT nor its expected behaviour.
\par
In this paper we study the SIT phase diagram of the (111) oriented interface between the two band insulators \LAO~and \STO. The interface has a gate tunable carrier density and it can form on the various faces of the \STO~crystal: (100), (110) and (111) \cite{herranz2012high}. While for (100) the cubic symmetry is projected onto the interface creating a square lattice, the (111) oriented \LAO$/$\STO~interface has a 2D triangular structure. This 2D crystalline symmetry is also reflected in the magneto-transport properties \cite{rout2017six}. Previous studies of this system found 2D superconductivity \cite{monteiro2017two,DavisMagnetoresistance2017} and a correlation between superconductivity and spin-orbit interaction \cite{rout2017link}.
\par
We use gate bias to tune the sample from the metallic and superconducting regime to the insulating regime. At various gate voltages we study the magnetic field response for parallel and perpendicular field orientations. We observe giant magnetoresistance features similar to those observed in amorphous 2D superconductors \cite{Steiner2008Approach,KapitulnikInOxPeak,ShaharInOxPeak,Baturina2007Quantum,BaturinaTiN} previously unseen in a crystalline material.
From the comparison between the effects of parallel and perpendicular magnetic fields we define an energy scale for the suppression of the insulating state via the breaking of Cooper pairs. This anisotropic magnetoresistance as well as the linear magnetoresistance observed at low fields and the hysteresis of the magnetoresistance features show that vortex excitations are responsible for the SIT. Surprisingly, these effects persist deep into the insulating state, revealing the importance of vortex excitations even when superconductivity is completely suppressed.

\begin{figure}
\begin{center}
\includegraphics[width=1\hsize]{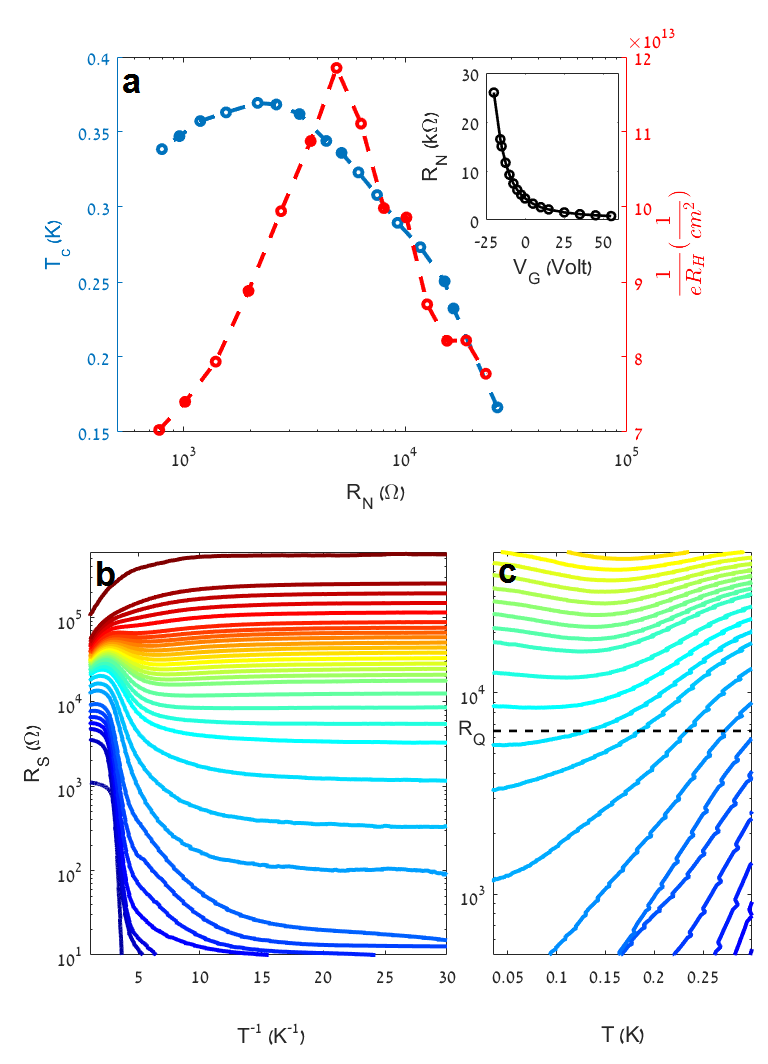}
\caption{(a) The superconducting critical temperature \Tc~(defined as the temperature for which the resistance drops to half its value at 1 K) and the inverse hall coefficient $\frac{1}{eR_H}$ measured at 2K are plotted as a function of the normal state sheet resistance $R_N$. The inset shows $R_N$ plotted as a function of the gate voltage $V_G$ in a particular cooldown. (b) Sheet resistance plotted in the logarithmic scale against $T^{-1}$ for gate voltages ranging from 30 (dark blue) to -190 (dark red) Volts. (c) Sheet resistance plotted against $T$ for those same voltages near the critical point, where $\frac{dR_S}{dT}_{T\rightarrow0}=0$. The black dashed line indicates the value of the quantum resistance $R_Q=\frac{h}{4e^2}$.}
\label{Fig-DomesAndGates}
\end{center}
\end{figure}

\bigskip
\noindent\textbf{Results}

\noindent\textbf{Effect of electrostatic gating.} The large dielectric constant of \STO, of the order of 20000 times that of the vacuum, allows us to strongly modulate the carrier density at the interface with a back-gate voltage \cite{caviglia2008electric}. We show the sheet resistance as a function of gate voltage at 1K in the inset of Fig.~\ref{Fig-DomesAndGates} (a). Because the response to gate voltage changes between one cool-down and the other (due to different domain configuration \cite{kalisky2013locally,IlaniSTOLAODomains} and trapped charges \cite{Biscaras2014}) we plot all sample properties as a function of the normal state sheet resistance $R_N$ rather than V$_g$.
\par
The superconducting transition temperature \Tc~and the low field inverse Hall coefficient $\frac{1}{eR_H}$ are plotted versus R$_N$ in Fig.~\ref{Fig-DomesAndGates} (a). The low resistance regime is consistent with Ref. \cite{rout2017link,davis2017anisotropic}. In the high resistance region the effect of the gate voltage on $\frac{1}{eR_H}$ reverses. This unusual behavior can be attributed to the combined contributions of holes and electrons \cite{STOLAO111Holes}, (which is consistent with the polar structure of the interface \cite{rout2017link}) and to electronic correlations \cite{maniv2015strong}. The latter of which has been invoked to explain a similar but weaker effect observed in (100) interfaces, where holes are not expected.
\par
The response of $\frac{1}{eR_H}$ to V$_g$ is much stronger than that seen for the (100) \STO/\LAO~interface \cite{bellDominant}. However, the change in carrier density is still not enough to explain the large change in resistance with V$_g$, implying that the mobility is the dominant factor in the gate dependence of the sheet resistance R$_S$, similar to what was suggested for the (100) interface \cite{bellDominant}. The effect of gate voltage on superconductivity is therefore twofold: first, it changes electron density and hence superfluid stiffness. Second, it modifies the effective disorder in the electron liquid, possibly by bringing the liquid closer to the interface. We therefore study the SIT both as a function of gate voltage and magnetic field.

\medskip
\noindent\textbf{Superconductor to insulator transition induced by electrostatic gating.}
For positive V$_g$, R$_S$ goes to zero within error at low temperatures (Fig.~\ref{Fig-DomesAndGates} (b)). As V$_g$ is decreased, the sample transitions into an anomalous metallic state with finite resistance at zero temperature \cite{kapitulnik2017anomalousReview}.  At the more negative gate voltages the sample transitions to an insulating state, for which $\frac{dR}{dT}<0$ in the measured temperature range. As shown in Fig.~\ref{Fig-DomesAndGates} (c), the resistance as a function of temperature flattens as the sheet resistance reaches R$_Q = h/4e^2$. These results are similar to those of Haviland and Goldman \cite{HavilandandGoldmannSITthickness} where the thickness of the film was the control parameter for the quantum phase transition  rather than V$_g$.

\begin{figure*}
\begin{center}
\includegraphics[width=1\hsize]{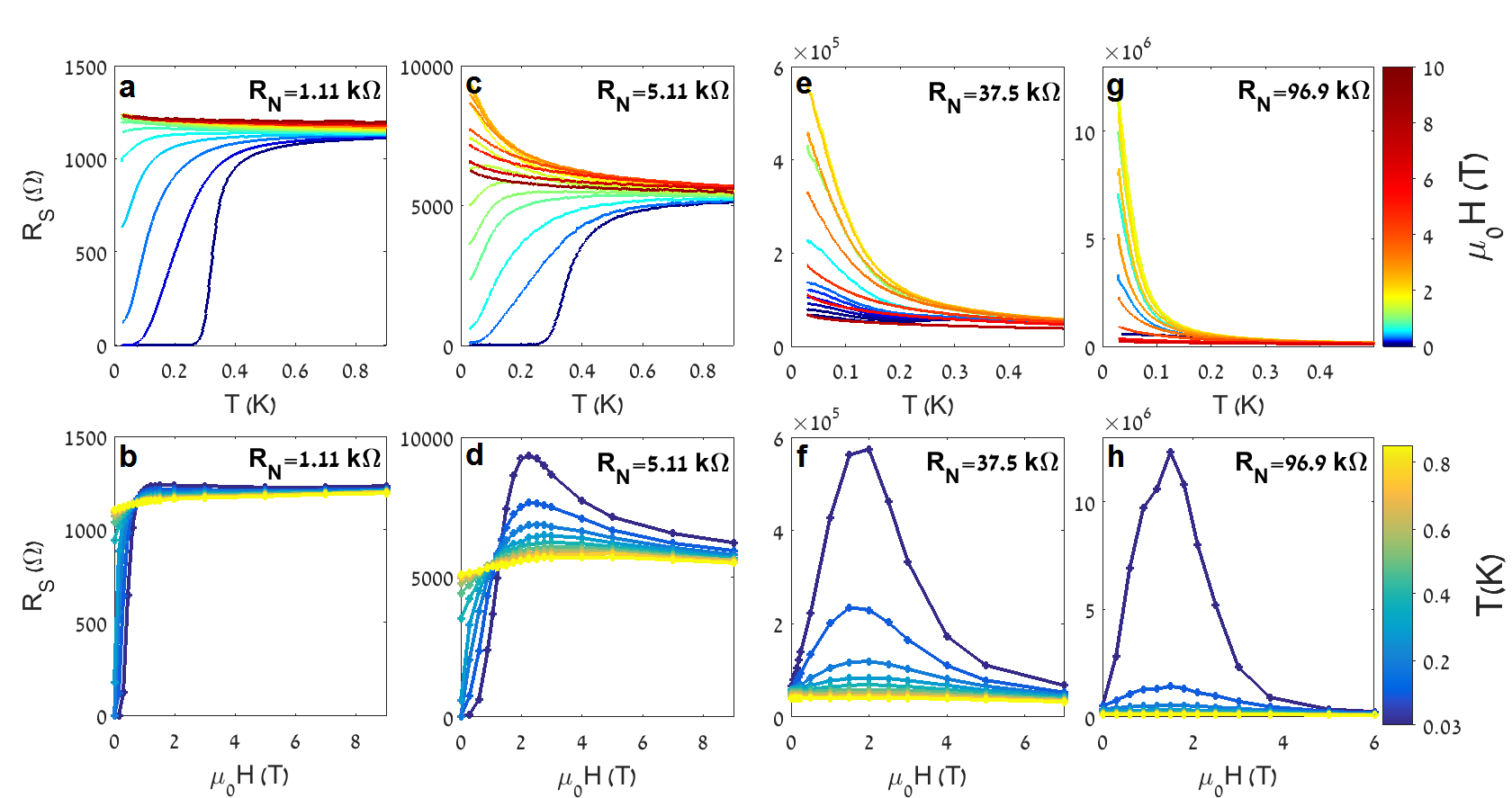}
\caption{(a),(c),(e) and (g)  Sheet resistance plotted against temperature with applied magnetic fields ranging from 0 T (dark blue) to 9 T (dark red) for four different gate voltages (labeled according to their R$_N$). (b),(d),(f) and (g) Sheet resistance plotted against magnetic field at different temperatures ranging from 0.035 K (dark blue) to 0.85 K (yellow) for the same gate voltages.}
\label{Fig-RTsRHs}
\end{center}
\end{figure*}

\medskip
\noindent\textbf{Superconductor to insulator transition induced by perpendicular magnetic field.}
In Fig.~\ref{Fig-RTsRHs} we present the perpendicular magnetic field induced SIT in four different regimes characterized by their R$_N$ values. For R$_N$ = 1.11 k$\Omega$ the sample is superconducting at low magnetic fields, but as the field is increased it transitions into a weakly insulating state. When R$_N$ is increased to 5.11 k$\Omega$, the sample is still superconducting and transitions to an insulating state under the application of magnetic field. However, at some magnetic field R$_S$ reaches a maximum value and further increase of the field destroys this insulating behaviour. For R$_N$ = 37.5 k$\Omega$, R$_S$ remains finite at low fields, yet the response to magnetic field becomes significantly stronger. Upon increasing R$_N$ even further to 96.9 k$\Omega$ the zero field R$_S$(T) is insulating-like with no clear signature of superconductivity, yet the relative amplitude of the magnetoresistance peak is larger than that of the previous gate voltages. A peak in the magnetoresistance has been previously shown to be related to the SIT in amorphous superconductors \cite{Steiner2008Approach,KapitulnikInOxPeak,ShaharInOxPeak,Baturina2007Quantum,BaturinaTiN} but in these experiments the different regimes could only be accessed in different samples. In our measurements, the continuous evolution of the magnetoresistance peak from the low R$_N$ to the high R$_N$ shows that even when the sample is insulating at zero field, the mechanism responsible for the SIT still effects the transport properties of the system.

In Fig.~\ref{Fig-DomesAndGates} (c) we showed that when changing V$_g$ at zero field, the value of R$_S$ at the SIT is very close to R$_Q$, as is expected from the self duality between Cooper pairs and vortices in 2D superconductors \cite{Fisher1990Presence}. However, we do not observe any universal resistance value in the field induced SIT. The lack of a singular critical point in isotherms in Figs.~\ref{Fig-RTsRHs} (b),(d),(f) and (h) as well as the nonomonotic behaviour in Figs.~\ref{Fig-DomesAndGates} (b) and \ref{Fig-RTsRHs} (e) can be the result of parallel fermionic channels (similar to the results seen by  Goldman \etal  \cite{GoldmanVariedSIT}).We deduce that these inhomogeneities play a less significant role at zero field from the fact that the exact quantum resistance is observed in the gate induced SIT.

Previous SIT experiments show thermally activated Arrhenius transport, an experimental feature of Bose insulator \cite{ShaharInOxPeak,KapitulnikInOxPeak,BaturinaTiN,2eMROscillations} resulting from the emergence of self-induced inhomogeneity leading to superconducting islands embedded in an insulating media in a uniformly disordered film \cite{KapitulnikShimshoniQuantumMelts,dubi2007nature,Skvortsov2005Superconductivity,ShaharIslandsSTM}. The electrons created by thermally breaking Cooper pairs take part in inter-island tunneling processes leading to Arrhenius transport with activation temperatures of the order of \Tc  \cite{ShaharInOxPeak,BaturinaTiN}. In the supplementary we present Arrhenius plots for the resistance at the magnetoresistance peak $H_{peak}$ in different gate voltages. While the data follow the correct behaviour at high temperatures (similar to other experiments \cite{KapitulnikPhysicaC}), the calculated values of the activation energy $T_A$ are lower than \Tc, implying inter-island tunneling is not the main mechanism responsible for the insulating state.
\begin{figure*}
\begin{center}
\includegraphics[width=1\hsize]{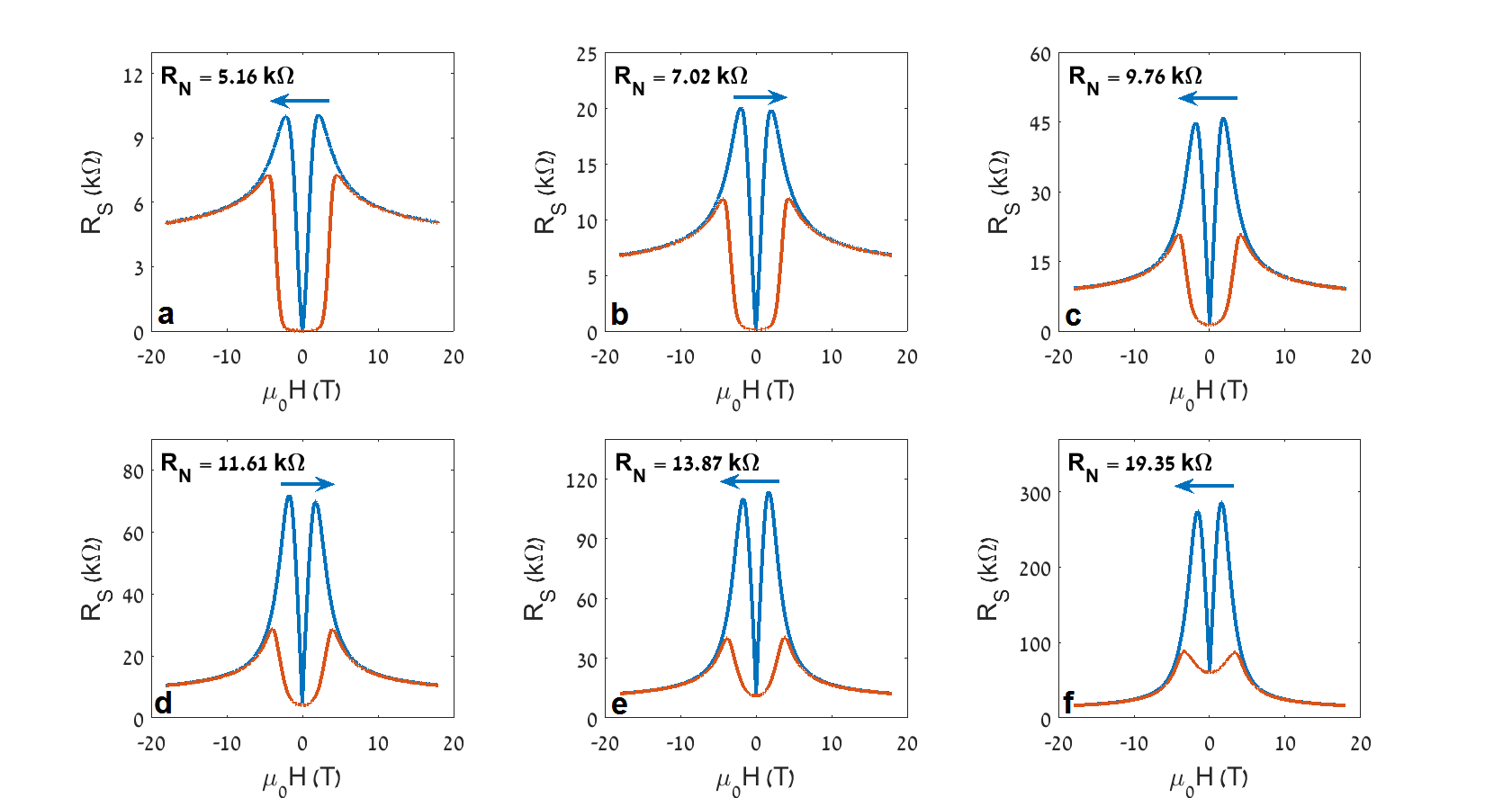}
\caption{Sheet resistance plotted against magnetic fields perpendicular (blue) and parallel (red) to the sample surface for six different gate voltages (labeled according to their R$_N$) at base temperature. The small asymmetry in the magnetoresistance peaks observed for perpendicular field orientation is related to the sweep direction (indicated by the blue arrows).}
\label{Fig-Tallahassee}
\end{center}
\end{figure*}

\begin{figure*}
\begin{center}
\includegraphics[width=1\hsize]{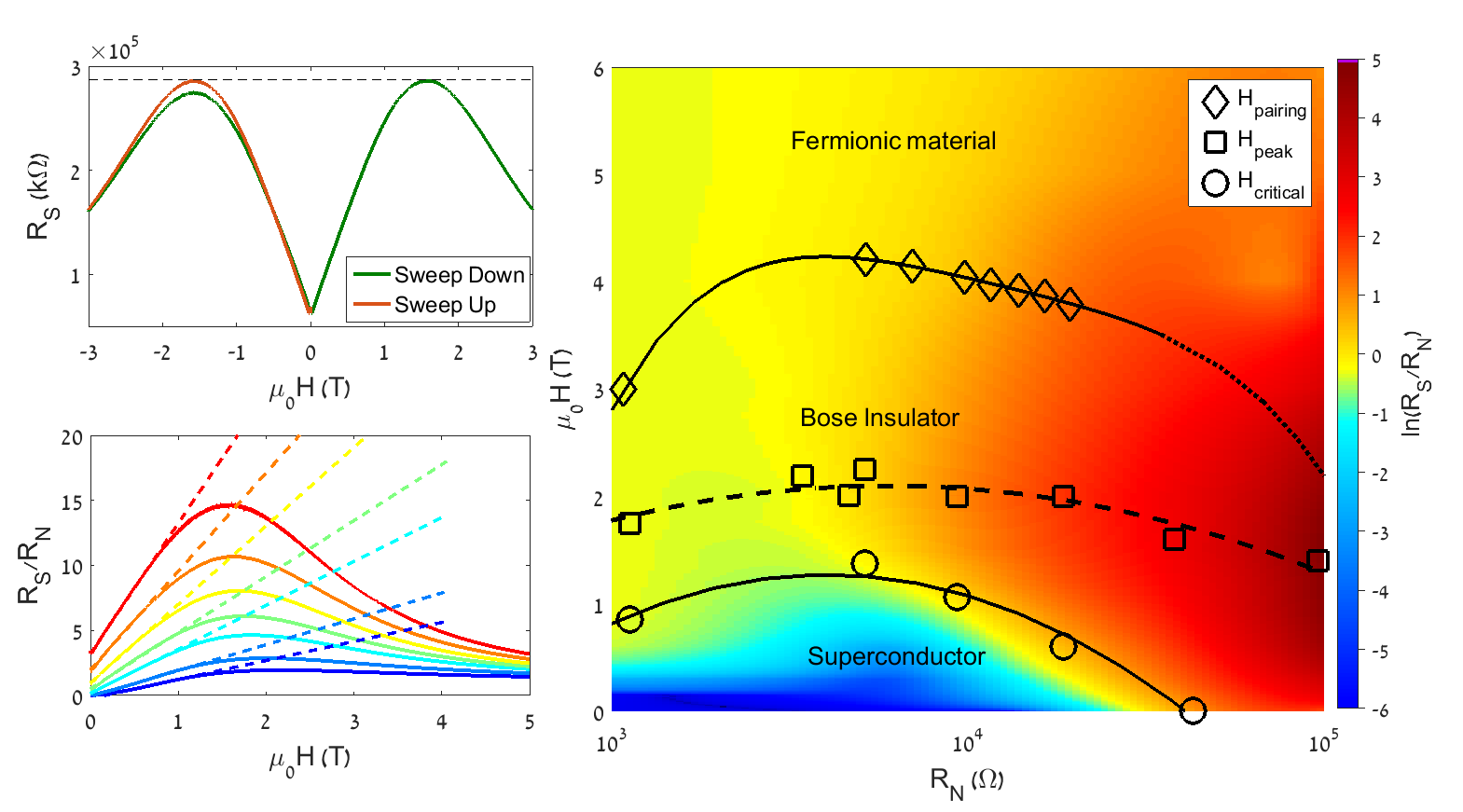}
\caption{(a) Sheet resistance as a function of magnetic field measured while sweeping the field down (red) and up (green) at base temperature. Both positive and negative field peaks have the same resistance value while sweeping towards the zero, indicating that the peak height is different for increasing or decreasing $|H|$ (away or towards zero). This hysteresis behaviour is different from that observed in systems with magnetic order, e.g. \STO/\LAO~nanowires \cite{ron2014anomalous}. (b)  Sheet resistance normalized by the normal state resistance plotted as a function of magnetic field for R$_N$ from 5.16 $k\Omega$ (blue) to 19.35 $k\Omega$ (red) at base temperature. The dashed lines show low field regions fits to the flux flow behavior: R$_S$/R$_N$ $\propto$ H. (c) Ln(R$_S$/R$_N$) plotted against R$_N$ and H at base temperature. The SIT critical field H$_{critical}$ defined as the field for which $\frac{dR_S}{dT}_{T\rightarrow0}=0$, the field H$_{peak}$ corresponding to magnetoresistance peak and the depairing field H$_{pairing}$ are also plotted as a function of normal state resistances, with the black lines as a guide to the eye (see Supplementary Information for more details). The solid lines show the borders between the different states: the superconducting region below H$_{critical}$, the Bose insulator between H$_{critical}$ and H$_{pairing}$ and the Fermionic material beyond H$_{pairing}$. The dotted part of the top line marks that the border is an extrapolation in these resistance values. The dashed line shows the points where the effect of the field on resistance changes direction.}
\label{Fig-ProVortexPlots}
\end{center}
\end{figure*}

\medskip
\noindent\textbf{Response to different magnetic field orientations.}
To further understand the origin of the SIT, we performed magnetic field sweeps in parallel and perpendicular orientations (see Fig.~\ref{Fig-Tallahassee}). First, we note that the high field magnetoresistance is isotropic, suggesting the absence of orbital effects in the normal state. This is contrasted with the highly anisotropic behavior at lower fields, where fluctuations can still survive. The observed anisotropy supports the idea that the effect seen under perpendicular fields is caused by vortex excitations.

Furthermore, we note that the magnetoresistance peak is hysteretic for increasing and decreasing fields (as seen in Figs. \ref{Fig-Tallahassee} and \ref{Fig-ProVortexPlots} (a)). In the vortex picture, this hysteresis is related to the effect of pinning potentials in the vortex regime \cite{VortexPinningHysteresis}. When the field is swept down from higher values the vortices move more freely resulting in more dissipation and a larger resistance at the peak. In the parallel field orientation, where no field induced vortices exist, the observed hysteresis is much smaller.

The linear field dependence of R$_S$ for small perpendicular H [Fig.~\ref{Fig-ProVortexPlots} (b)] is similar to flux-flow-type behaviour, where as the field is increased more vortices are created and the resistance increases. Surprisingly, even for negative gate voltages where R$_S$ at zero field is $50 k\Omega$ the linear in field behavior is still observed. This implies that mobile vortex excitations exist even for such high sheet resistance.

In comparison to the perpendicular field response, the response to parallel fields at low R$_N$ starts with a flat region with zero resistance until at some magnetic field the resistance increases sharply and eventually converges with the perpendicular field curve at H$_{pairing}$. We interpret H$_{pairing}$ as the Zeeman depairing field. As R$_N$ is increased the zero field superconductivity fails \cite{kapitulnik2017anomalousReview} but H$_{pairing}$ can be easily identified.
The qualitative behavior of the magnetoresistance at intermediate parallel magnetic fields is consistent with the picture of superconducting fluctuations destroyed by the magnetic field.

\bigskip
\noindent\textbf{Discussion}

Fig.~\ref{Fig-ProVortexPlots} (c) shows the R$_N$-H phase diagram of our interface revealing three different states of the system, separated by the superconductor to insulator critical field $H_{critical}$ and the depairing field H$_{pairing}$. H$_{critical}$ marks the transition from the regime where condensed Cooper pairs dominate the transport, to an insulating phase. This field gradually vanishes with increasing R$_N$ similar to previous SIT studies on \STO/\LAO~interfaces \cite{Lesueur2013multiplecriticality,Mehta2014Magnetic,GriffithsSTOLAO}.

Our results indicate that that this insulating phase is a Bose-condensed liquid of vortex excitations, a consequence of the duality between Cooper pairs and vortices \cite{FisherSITTheory}. The existence of pairing in the insulating side of SIT has been seen from magnetoresistance oscillations with 2$e$ period for a nanohoneycomb array of holes in Bi film  \cite{2eMROscillations} and the direct measurement of the superconducting gap \cite{ShaharIslandsJunction,ShaharIslandsSTM}. The observed large magnetorsistance peak is one of the signatures of such Bose insulator \cite{Steiner2008Approach,KapitulnikInOxPeak,ShaharInOxPeak,Baturina2007Quantum,BaturinaTiN}. Surprisingly, our results show that the magnetoresistance peak persists well beyond the point where the upper critical field drops to zero, that is, when the sample is insulating at zero field.

At the new field scale that we define, H$_{pairing}$, the Zeeman energy exceeds the pairing one and the material transitions from the Bose-insulator to the fermionic material, where no Cooper pairs exist and fermionic physics determines the transport properties.
\par
Following Emery and Kivelson \cite{KivelsonEmery1995} we estimate the temperature for which superconducting phase order disappears. We find that this temperature is of the order of 10 Kelvin (see Supplementary Information for more details) which is roughly of the order of $\mu_{B}H_{depairing}/k_B$. This observation reinforces our finding that the vortex excitations are an important factor in the insulating phase and its field response. The observed H$_{pairing}$ agrees quite well with the values for the (100) \STO/\LAO~ quantum dot \cite{Levy2015QD}. In contrast, the tunneling spectra for the (100) interface revealed that the pseudogap-like features vanish at a much lower energies \cite{Manhart2013Tunneling}.

The phase diagram resembles that of 2D amorphous superconductors, however, our material is a 2D crystalline heterostructure devoid of any structural inhomogeneity or granularity. Furthermore, the dielectric constant of \STO~is so large one would expect local disorder potential to be screened, rendering the system cleaner. We also note that while the magnetic field induced SIT has been reported for (100) and (110) \STO/LaTiO$_3$ interfaces \cite{Lesueur2013multiplecriticality,Mehta2014Magnetic,GriffithsSTOLAO}, a clear magnetoresistance peak was not observed and the nature of highly insulating regime has not been investigated.
We speculate that the more extreme SIT observed in the (111) interface compared to the (100) and (110) cases may be related to frustrated antiferromagnetic coupling in the (111) triangular interface, which allows superconductivity to exist in a broader R$_N$ region. Such antiferromagnetic coupling has been observed in (100) \STO/\LAO~nanowires \cite{ron2014anomalous}.

In summary, we study the superconductor-to-insulator transition in a (111) \STO/\LAO~interface. The quantum phase transition is controlled by gate voltage and magnetic field. The tunability of our system allows us to follow features related to the superconductor-to-insulator transition such as the magnetoresistance peak and the depairing field H$_{pairing}$ deep into the insulating state. We observe a gate-controlled transition from the superconducting to the insulating state at the quantum resistance similar to the hallmark data of Haviland and Goldman \cite{HavilandandGoldmannSITthickness}. We use the comparison of measurements in parallel and perpendicular field to define and follow a new energy scale related to the depairing field H$_{pairing}$. The linear field dependence of the magnetoresistance at low fields, the strong anisotropy of the magnetoresistance at the peak region and the hysteresis of the peak all are evidence that this peak is related to the formation of a liquid of vortex excitations. Our data provide a broad view on the evolution of a variety of phenomena observed, until now, in many different samples and regimes and show that vortices play an important role in the insulating state observed beyond the superconductor-to-insulator transition.

\bigskip
\noindent\textbf{Methods}

\noindent\textbf{Sample fabrication.} Epitaxial films of LaAlO$_3$ were deposited on atomically flat \STO~(111) substrate using pulsed laser deposition. The details of deposition procedure and substrate treatment are described in Ref. \cite{rout2017six,rout2017link}. We monitor the layer-by-layer growth of 14 monolayers (LaO$_3$/Al layers) by reflection high energy electron diffraction (RHEED) oscillations (see Supplementary Information for more details). Atomic force microscope images show the step and terrace morphology of the film with step heights of 0.22 nm. A gold back-gate electrode is evaporated on the bottom of the \STO crystal. The positive voltage terminal is connected to the bottom gate electrode.

\textbf{Transport measurements.} The sample was measured in an Oxford Instruments Triton 400 dilution refrigerator (with a base temperature of $\sim$30 mK) and in a wet dilution refrigerator at the National High Magnetic Field Laboratory (NHMFL) in Tallahassee (with a base temperature of $\sim$20 mK). Hall measurements were performed in a He$^4$ cryostat at a temperature of 1.5 K. The measurements were conducted with a Keithley Current Source and Nanovoltmeter in a 4-point configuration. IV measurements were taken in order to make sure the currents used in resistance measurements are in the linear response regime. The currents used were between 5$\times 10^{-10}$ to 1$\times 10^{-8}$ Ampere.

\bibliographystyle{apsrev}
\bibliography{paper_bib}

\begin{thebibliography}{49}
\expandafter\ifx\csname natexlab\endcsname\relax\def\natexlab#1{#1}\fi
\expandafter\ifx\csname bibnamefont\endcsname\relax
  \def\bibnamefont#1{#1}\fi
\expandafter\ifx\csname bibfnamefont\endcsname\relax
  \def\bibfnamefont#1{#1}\fi
\expandafter\ifx\csname citenamefont\endcsname\relax
  \def\citenamefont#1{#1}\fi
\expandafter\ifx\csname url\endcsname\relax
  \def\url#1{\texttt{#1}}\fi
\expandafter\ifx\csname urlprefix\endcsname\relax\def\urlprefix{URL }\fi
\providecommand{\bibinfo}[2]{#2}
\providecommand{\eprint}[2][]{\url{#2}}

\bibitem[{\citenamefont{Haviland et~al.}(1989)\citenamefont{Haviland, Liu, and
  Goldman}}]{HavilandandGoldmannSITthickness}
\bibinfo{author}{\bibfnamefont{D.~B.} \bibnamefont{Haviland}},
  \bibinfo{author}{\bibfnamefont{Y.}~\bibnamefont{Liu}}, \bibnamefont{and}
  \bibinfo{author}{\bibfnamefont{A.~M.} \bibnamefont{Goldman}},
  \bibinfo{journal}{Phys. Rev. Lett.} \textbf{\bibinfo{volume}{62}},
  \bibinfo{pages}{2180} (\bibinfo{year}{1989}),
  \urlprefix\url{https://link.aps.org/doi/10.1103/PhysRevLett.62.2180}.

\bibitem[{\citenamefont{Hebard and Paalanen}(1990)}]{HebardPaalanenInOx}
\bibinfo{author}{\bibfnamefont{A.~F.} \bibnamefont{Hebard}} \bibnamefont{and}
  \bibinfo{author}{\bibfnamefont{M.~A.} \bibnamefont{Paalanen}},
  \bibinfo{journal}{Phys. Rev. Lett.} \textbf{\bibinfo{volume}{65}},
  \bibinfo{pages}{927} (\bibinfo{year}{1990}),
  \urlprefix\url{https://link.aps.org/doi/10.1103/PhysRevLett.65.927}.

\bibitem[{\citenamefont{Sambandamurthy
  et~al.}(2004)\citenamefont{Sambandamurthy, Engel, Johansson, and
  Shahar}}]{ShaharInOxPeak}
\bibinfo{author}{\bibfnamefont{G.}~\bibnamefont{Sambandamurthy}},
  \bibinfo{author}{\bibfnamefont{L.~W.} \bibnamefont{Engel}},
  \bibinfo{author}{\bibfnamefont{A.}~\bibnamefont{Johansson}},
  \bibnamefont{and} \bibinfo{author}{\bibfnamefont{D.}~\bibnamefont{Shahar}},
  \bibinfo{journal}{Phys. Rev. Lett.} \textbf{\bibinfo{volume}{92}},
  \bibinfo{pages}{107005} (\bibinfo{year}{2004}),
  \urlprefix\url{https://link.aps.org/doi/10.1103/PhysRevLett.92.107005}.

\bibitem[{\citenamefont{Ovadia et~al.}(2013)\citenamefont{Ovadia, Kalok,
  Sac{\'e}p{\'e}, and Shahar}}]{ovadia2013duality}
\bibinfo{author}{\bibfnamefont{M.}~\bibnamefont{Ovadia}},
  \bibinfo{author}{\bibfnamefont{D.}~\bibnamefont{Kalok}},
  \bibinfo{author}{\bibfnamefont{B.}~\bibnamefont{Sac{\'e}p{\'e}}},
  \bibnamefont{and} \bibinfo{author}{\bibfnamefont{D.}~\bibnamefont{Shahar}},
  \bibinfo{journal}{Nature Physics} \textbf{\bibinfo{volume}{9}},
  \bibinfo{pages}{415} (\bibinfo{year}{2013}).

\bibitem[{\citenamefont{Steiner et~al.}(2008)\citenamefont{Steiner, Breznay,
  and Kapitulnik}}]{Steiner2008Approach}
\bibinfo{author}{\bibfnamefont{M.~A.} \bibnamefont{Steiner}},
  \bibinfo{author}{\bibfnamefont{N.~P.} \bibnamefont{Breznay}},
  \bibnamefont{and}
  \bibinfo{author}{\bibfnamefont{A.}~\bibnamefont{Kapitulnik}},
  \bibinfo{journal}{Phys. Rev. B} \textbf{\bibinfo{volume}{77}},
  \bibinfo{pages}{212501} (\bibinfo{year}{2008}),
  \urlprefix\url{https://link.aps.org/doi/10.1103/PhysRevB.77.212501}.

\bibitem[{\citenamefont{Steiner et~al.}(2005)\citenamefont{Steiner, Boebinger,
  and Kapitulnik}}]{KapitulnikInOxPeak}
\bibinfo{author}{\bibfnamefont{M.~A.} \bibnamefont{Steiner}},
  \bibinfo{author}{\bibfnamefont{G.}~\bibnamefont{Boebinger}},
  \bibnamefont{and}
  \bibinfo{author}{\bibfnamefont{A.}~\bibnamefont{Kapitulnik}},
  \bibinfo{journal}{Phys. Rev. Lett.} \textbf{\bibinfo{volume}{94}},
  \bibinfo{pages}{107008} (\bibinfo{year}{2005}),
  \urlprefix\url{https://link.aps.org/doi/10.1103/PhysRevLett.94.107008}.

\bibitem[{\citenamefont{Yazdani and Kapitulnik}(1995)}]{KapitulnikMoGe}
\bibinfo{author}{\bibfnamefont{A.}~\bibnamefont{Yazdani}} \bibnamefont{and}
  \bibinfo{author}{\bibfnamefont{A.}~\bibnamefont{Kapitulnik}},
  \bibinfo{journal}{Phys. Rev. Lett.} \textbf{\bibinfo{volume}{74}},
  \bibinfo{pages}{3037} (\bibinfo{year}{1995}),
  \urlprefix\url{https://link.aps.org/doi/10.1103/PhysRevLett.74.3037}.

\bibitem[{\citenamefont{Baturina
  et~al.}(2007{\natexlab{a}})\citenamefont{Baturina, Strunk, Baklanov, and
  Satta}}]{Baturina2007Quantum}
\bibinfo{author}{\bibfnamefont{T.~I.} \bibnamefont{Baturina}},
  \bibinfo{author}{\bibfnamefont{C.}~\bibnamefont{Strunk}},
  \bibinfo{author}{\bibfnamefont{M.~R.} \bibnamefont{Baklanov}},
  \bibnamefont{and} \bibinfo{author}{\bibfnamefont{A.}~\bibnamefont{Satta}},
  \bibinfo{journal}{Phys. Rev. Lett.} \textbf{\bibinfo{volume}{98}},
  \bibinfo{pages}{127003} (\bibinfo{year}{2007}{\natexlab{a}}),
  \urlprefix\url{https://link.aps.org/doi/10.1103/PhysRevLett.98.127003}.

\bibitem[{\citenamefont{Baturina
  et~al.}(2007{\natexlab{b}})\citenamefont{Baturina, Mironov, Vinokur,
  Baklanov, and Strunk}}]{BaturinaTiN}
\bibinfo{author}{\bibfnamefont{T.~I.} \bibnamefont{Baturina}},
  \bibinfo{author}{\bibfnamefont{A.~Y.} \bibnamefont{Mironov}},
  \bibinfo{author}{\bibfnamefont{V.~M.} \bibnamefont{Vinokur}},
  \bibinfo{author}{\bibfnamefont{M.~R.} \bibnamefont{Baklanov}},
  \bibnamefont{and} \bibinfo{author}{\bibfnamefont{C.}~\bibnamefont{Strunk}},
  \bibinfo{journal}{Phys. Rev. Lett.} \textbf{\bibinfo{volume}{99}},
  \bibinfo{pages}{257003} (\bibinfo{year}{2007}{\natexlab{b}}),
  \urlprefix\url{https://link.aps.org/doi/10.1103/PhysRevLett.99.257003}.

\bibitem[{\citenamefont{Leng et~al.}(2011)\citenamefont{Leng,
  Garcia-Barriocanal, Bose, Lee, and Goldman}}]{Leng2011Electrostatic}
\bibinfo{author}{\bibfnamefont{X.}~\bibnamefont{Leng}},
  \bibinfo{author}{\bibfnamefont{J.}~\bibnamefont{Garcia-Barriocanal}},
  \bibinfo{author}{\bibfnamefont{S.}~\bibnamefont{Bose}},
  \bibinfo{author}{\bibfnamefont{Y.}~\bibnamefont{Lee}}, \bibnamefont{and}
  \bibinfo{author}{\bibfnamefont{A.~M.} \bibnamefont{Goldman}},
  \bibinfo{journal}{Phys. Rev. Lett.} \textbf{\bibinfo{volume}{107}},
  \bibinfo{pages}{027001} (\bibinfo{year}{2011}),
  \urlprefix\url{https://link.aps.org/doi/10.1103/PhysRevLett.107.027001}.

\bibitem[{\citenamefont{Bollinger et~al.}(2011)\citenamefont{Bollinger, Dubuis,
  Yoon, Pavuna, Misewich, and Bo{\v{z}}ovi{\'c}}}]{BozovicLSCO}
\bibinfo{author}{\bibfnamefont{A.~T.} \bibnamefont{Bollinger}},
  \bibinfo{author}{\bibfnamefont{G.}~\bibnamefont{Dubuis}},
  \bibinfo{author}{\bibfnamefont{J.}~\bibnamefont{Yoon}},
  \bibinfo{author}{\bibfnamefont{D.}~\bibnamefont{Pavuna}},
  \bibinfo{author}{\bibfnamefont{J.}~\bibnamefont{Misewich}}, \bibnamefont{and}
  \bibinfo{author}{\bibfnamefont{I.}~\bibnamefont{Bo{\v{z}}ovi{\'c}}},
  \bibinfo{journal}{Nature} \textbf{\bibinfo{volume}{472}},
  \bibinfo{pages}{458} (\bibinfo{year}{2011}).

\bibitem[{\citenamefont{Jaeger et~al.}(1989)\citenamefont{Jaeger, Haviland,
  Orr, and Goldman}}]{GoldmanVariedSIT}
\bibinfo{author}{\bibfnamefont{H.~M.} \bibnamefont{Jaeger}},
  \bibinfo{author}{\bibfnamefont{D.~B.} \bibnamefont{Haviland}},
  \bibinfo{author}{\bibfnamefont{B.~G.} \bibnamefont{Orr}}, \bibnamefont{and}
  \bibinfo{author}{\bibfnamefont{A.~M.} \bibnamefont{Goldman}},
  \bibinfo{journal}{Phys. Rev. B} \textbf{\bibinfo{volume}{40}},
  \bibinfo{pages}{182} (\bibinfo{year}{1989}),
  \urlprefix\url{https://link.aps.org/doi/10.1103/PhysRevB.40.182}.

\bibitem[{\citenamefont{Qin et~al.}(2006)\citenamefont{Qin, Vicente, and
  Yoon}}]{SITTantalum}
\bibinfo{author}{\bibfnamefont{Y.}~\bibnamefont{Qin}},
  \bibinfo{author}{\bibfnamefont{C.~L.} \bibnamefont{Vicente}},
  \bibnamefont{and} \bibinfo{author}{\bibfnamefont{J.}~\bibnamefont{Yoon}},
  \bibinfo{journal}{Phys. Rev. B} \textbf{\bibinfo{volume}{73}},
  \bibinfo{pages}{100505} (\bibinfo{year}{2006}),
  \urlprefix\url{https://link.aps.org/doi/10.1103/PhysRevB.73.100505}.

\bibitem[{\citenamefont{Aubin et~al.}(2006)\citenamefont{Aubin,
  Marrache-Kikuchi, Pourret, Behnia, Berg\'e, Dumoulin, and Lesueur}}]{SITNbSi}
\bibinfo{author}{\bibfnamefont{H.}~\bibnamefont{Aubin}},
  \bibinfo{author}{\bibfnamefont{C.~A.} \bibnamefont{Marrache-Kikuchi}},
  \bibinfo{author}{\bibfnamefont{A.}~\bibnamefont{Pourret}},
  \bibinfo{author}{\bibfnamefont{K.}~\bibnamefont{Behnia}},
  \bibinfo{author}{\bibfnamefont{L.}~\bibnamefont{Berg\'e}},
  \bibinfo{author}{\bibfnamefont{L.}~\bibnamefont{Dumoulin}}, \bibnamefont{and}
  \bibinfo{author}{\bibfnamefont{J.}~\bibnamefont{Lesueur}},
  \bibinfo{journal}{Phys. Rev. B} \textbf{\bibinfo{volume}{73}},
  \bibinfo{pages}{094521} (\bibinfo{year}{2006}),
  \urlprefix\url{https://link.aps.org/doi/10.1103/PhysRevB.73.094521}.

\bibitem[{\citenamefont{Stewart et~al.}(2007)\citenamefont{Stewart, Yin, Xu,
  and Valles}}]{2eMROscillations}
\bibinfo{author}{\bibfnamefont{M.~D.} \bibnamefont{Stewart}},
  \bibinfo{author}{\bibfnamefont{A.}~\bibnamefont{Yin}},
  \bibinfo{author}{\bibfnamefont{J.~M.} \bibnamefont{Xu}}, \bibnamefont{and}
  \bibinfo{author}{\bibfnamefont{J.~M.} \bibnamefont{Valles}},
  \bibinfo{journal}{Science} \textbf{\bibinfo{volume}{318}},
  \bibinfo{pages}{1273} (\bibinfo{year}{2007}), ISSN \bibinfo{issn}{0036-8075},
  \eprint{http://science.sciencemag.org/content/318/5854/1273.full.pdf},
  \urlprefix\url{http://science.sciencemag.org/content/318/5854/1273}.

\bibitem[{\citenamefont{White et~al.}(1986)\citenamefont{White, Dynes, and
  Garno}}]{SITSnAndPb}
\bibinfo{author}{\bibfnamefont{A.~E.} \bibnamefont{White}},
  \bibinfo{author}{\bibfnamefont{R.~C.} \bibnamefont{Dynes}}, \bibnamefont{and}
  \bibinfo{author}{\bibfnamefont{J.~P.} \bibnamefont{Garno}},
  \bibinfo{journal}{Phys. Rev. B} \textbf{\bibinfo{volume}{33}},
  \bibinfo{pages}{3549} (\bibinfo{year}{1986}),
  \urlprefix\url{https://link.aps.org/doi/10.1103/PhysRevB.33.3549}.

\bibitem[{\citenamefont{Finkel'stein}(1994)}]{finkel1994superconducting}
\bibinfo{author}{\bibfnamefont{A.}~\bibnamefont{Finkel'stein}},
  \bibinfo{journal}{Physica B: Condensed Matter}
  \textbf{\bibinfo{volume}{197}}, \bibinfo{pages}{636 } (\bibinfo{year}{1994}),
  ISSN \bibinfo{issn}{0921-4526},
  \urlprefix\url{http://www.sciencedirect.com/science/article/pii/0921452694902674}.

\bibitem[{\citenamefont{Burmistrov et~al.}(2015)\citenamefont{Burmistrov,
  Gornyi, and Mirlin}}]{SITTheoryMirlin}
\bibinfo{author}{\bibfnamefont{I.~S.} \bibnamefont{Burmistrov}},
  \bibinfo{author}{\bibfnamefont{I.~V.} \bibnamefont{Gornyi}},
  \bibnamefont{and} \bibinfo{author}{\bibfnamefont{A.~D.}
  \bibnamefont{Mirlin}}, \bibinfo{journal}{Phys. Rev. B}
  \textbf{\bibinfo{volume}{92}}, \bibinfo{pages}{014506}
  (\bibinfo{year}{2015}),
  \urlprefix\url{https://link.aps.org/doi/10.1103/PhysRevB.92.014506}.

\bibitem[{\citenamefont{Szab\'o et~al.}(2016)\citenamefont{Szab\'o, Samuely,
  Ha\ifmmode~\check{s}\else \v{s}\fi{}kov\'a, Ka\ifmmode \check{c}\else
  \v{c}\fi{}mar\ifmmode~\check{c}\else \v{c}\fi{}\'{\i}k, \ifmmode
  \check{Z}\else \v{Z}\fi{}emli\ifmmode~\check{c}\else \v{c}\fi{}ka, Grajcar,
  Rodrigo, and Samuely}}]{MoC_STM}
\bibinfo{author}{\bibfnamefont{P.}~\bibnamefont{Szab\'o}},
  \bibinfo{author}{\bibfnamefont{T.}~\bibnamefont{Samuely}},
  \bibinfo{author}{\bibfnamefont{V.}~\bibnamefont{Ha\ifmmode~\check{s}\else
  \v{s}\fi{}kov\'a}},
  \bibinfo{author}{\bibfnamefont{J.}~\bibnamefont{Ka\ifmmode \check{c}\else
  \v{c}\fi{}mar\ifmmode~\check{c}\else \v{c}\fi{}\'{\i}k}},
  \bibinfo{author}{\bibfnamefont{M.}~\bibnamefont{\ifmmode \check{Z}\else
  \v{Z}\fi{}emli\ifmmode~\check{c}\else \v{c}\fi{}ka}},
  \bibinfo{author}{\bibfnamefont{M.}~\bibnamefont{Grajcar}},
  \bibinfo{author}{\bibfnamefont{J.~G.} \bibnamefont{Rodrigo}},
  \bibnamefont{and} \bibinfo{author}{\bibfnamefont{P.}~\bibnamefont{Samuely}},
  \bibinfo{journal}{Phys. Rev. B} \textbf{\bibinfo{volume}{93}},
  \bibinfo{pages}{014505} (\bibinfo{year}{2016}),
  \urlprefix\url{https://link.aps.org/doi/10.1103/PhysRevB.93.014505}.

\bibitem[{\citenamefont{Shimshoni et~al.}(1998)\citenamefont{Shimshoni,
  Auerbach, and Kapitulnik}}]{KapitulnikShimshoniQuantumMelts}
\bibinfo{author}{\bibfnamefont{E.}~\bibnamefont{Shimshoni}},
  \bibinfo{author}{\bibfnamefont{A.}~\bibnamefont{Auerbach}}, \bibnamefont{and}
  \bibinfo{author}{\bibfnamefont{A.}~\bibnamefont{Kapitulnik}},
  \bibinfo{journal}{Phys. Rev. Lett.} \textbf{\bibinfo{volume}{80}},
  \bibinfo{pages}{3352} (\bibinfo{year}{1998}),
  \urlprefix\url{https://link.aps.org/doi/10.1103/PhysRevLett.80.3352}.

\bibitem[{\citenamefont{Dubi et~al.}(2007)\citenamefont{Dubi, Meir, and
  Avishai}}]{dubi2007nature}
\bibinfo{author}{\bibfnamefont{Y.}~\bibnamefont{Dubi}},
  \bibinfo{author}{\bibfnamefont{Y.}~\bibnamefont{Meir}}, \bibnamefont{and}
  \bibinfo{author}{\bibfnamefont{Y.}~\bibnamefont{Avishai}},
  \bibinfo{journal}{Nature} \textbf{\bibinfo{volume}{449}},
  \bibinfo{pages}{876} (\bibinfo{year}{2007}).

\bibitem[{\citenamefont{Skvortsov and
  Feigel'man}(2005)}]{Skvortsov2005Superconductivity}
\bibinfo{author}{\bibfnamefont{M.~A.} \bibnamefont{Skvortsov}}
  \bibnamefont{and} \bibinfo{author}{\bibfnamefont{M.~V.}
  \bibnamefont{Feigel'man}}, \bibinfo{journal}{Phys. Rev. Lett.}
  \textbf{\bibinfo{volume}{95}}, \bibinfo{pages}{057002}
  (\bibinfo{year}{2005}),
  \urlprefix\url{https://link.aps.org/doi/10.1103/PhysRevLett.95.057002}.

\bibitem[{\citenamefont{Fisher et~al.}(1990)\citenamefont{Fisher, Grinstein,
  and Girvin}}]{Fisher1990Presence}
\bibinfo{author}{\bibfnamefont{M.~P.~A.} \bibnamefont{Fisher}},
  \bibinfo{author}{\bibfnamefont{G.}~\bibnamefont{Grinstein}},
  \bibnamefont{and} \bibinfo{author}{\bibfnamefont{S.~M.}
  \bibnamefont{Girvin}}, \bibinfo{journal}{Phys. Rev. Lett.}
  \textbf{\bibinfo{volume}{64}}, \bibinfo{pages}{587} (\bibinfo{year}{1990}),
  \urlprefix\url{https://link.aps.org/doi/10.1103/PhysRevLett.64.587}.

\bibitem[{\citenamefont{Fisher}(1990)}]{FisherSITTheory}
\bibinfo{author}{\bibfnamefont{M.~P.~A.} \bibnamefont{Fisher}},
  \bibinfo{journal}{Phys. Rev. Lett.} \textbf{\bibinfo{volume}{65}},
  \bibinfo{pages}{923} (\bibinfo{year}{1990}),
  \urlprefix\url{https://link.aps.org/doi/10.1103/PhysRevLett.65.923}.

\bibitem[{\citenamefont{Herranz et~al.}(2012)\citenamefont{Herranz,
  S{\'a}nchez, Dix, Scigaj, and Fontcuberta}}]{herranz2012high}
\bibinfo{author}{\bibfnamefont{G.}~\bibnamefont{Herranz}},
  \bibinfo{author}{\bibfnamefont{F.}~\bibnamefont{S{\'a}nchez}},
  \bibinfo{author}{\bibfnamefont{N.}~\bibnamefont{Dix}},
  \bibinfo{author}{\bibfnamefont{M.}~\bibnamefont{Scigaj}}, \bibnamefont{and}
  \bibinfo{author}{\bibfnamefont{J.}~\bibnamefont{Fontcuberta}},
  \bibinfo{journal}{Sci. Rep.} \textbf{\bibinfo{volume}{2}},
  \bibinfo{pages}{758} (\bibinfo{year}{2012}).

\bibitem[{\citenamefont{Rout et~al.}(2017{\natexlab{a}})\citenamefont{Rout,
  Agireen, Maniv, Goldstein, and Dagan}}]{rout2017six}
\bibinfo{author}{\bibfnamefont{P.~K.} \bibnamefont{Rout}},
  \bibinfo{author}{\bibfnamefont{I.}~\bibnamefont{Agireen}},
  \bibinfo{author}{\bibfnamefont{E.}~\bibnamefont{Maniv}},
  \bibinfo{author}{\bibfnamefont{M.}~\bibnamefont{Goldstein}},
  \bibnamefont{and} \bibinfo{author}{\bibfnamefont{Y.}~\bibnamefont{Dagan}},
  \bibinfo{journal}{arXiv:1701.02153}  (\bibinfo{year}{2017}{\natexlab{a}}).

\bibitem[{\citenamefont{Monteiro et~al.}(2017)\citenamefont{Monteiro,
  Groenendijk, Groen, de~Bruijckere, Gaudenzi, van~der Zant, and
  Caviglia}}]{monteiro2017two}
\bibinfo{author}{\bibfnamefont{A.}~\bibnamefont{Monteiro}},
  \bibinfo{author}{\bibfnamefont{D.}~\bibnamefont{Groenendijk}},
  \bibinfo{author}{\bibfnamefont{I.}~\bibnamefont{Groen}},
  \bibinfo{author}{\bibfnamefont{J.}~\bibnamefont{de~Bruijckere}},
  \bibinfo{author}{\bibfnamefont{R.}~\bibnamefont{Gaudenzi}},
  \bibinfo{author}{\bibfnamefont{H.}~\bibnamefont{van~der Zant}},
  \bibnamefont{and} \bibinfo{author}{\bibfnamefont{A.}~\bibnamefont{Caviglia}},
  \bibinfo{journal}{arXiv:1703.04742}  (\bibinfo{year}{2017}).

\bibitem[{\citenamefont{Davis et~al.}(2017{\natexlab{a}})\citenamefont{Davis,
  Huang, Han, Ariando, Venkatesan, and
  Chandrasekhar}}]{DavisMagnetoresistance2017}
\bibinfo{author}{\bibfnamefont{S.}~\bibnamefont{Davis}},
  \bibinfo{author}{\bibfnamefont{Z.}~\bibnamefont{Huang}},
  \bibinfo{author}{\bibfnamefont{K.}~\bibnamefont{Han}},
  \bibinfo{author}{\bibnamefont{Ariando}},
  \bibinfo{author}{\bibfnamefont{T.}~\bibnamefont{Venkatesan}},
  \bibnamefont{and}
  \bibinfo{author}{\bibfnamefont{V.}~\bibnamefont{Chandrasekhar}},
  \bibinfo{journal}{Phys. Rev. B} \textbf{\bibinfo{volume}{96}},
  \bibinfo{pages}{134502} (\bibinfo{year}{2017}{\natexlab{a}}),
  \urlprefix\url{https://link.aps.org/doi/10.1103/PhysRevB.96.134502}.

\bibitem[{\citenamefont{Rout et~al.}(2017{\natexlab{b}})\citenamefont{Rout,
  Maniv, and Dagan}}]{rout2017link}
\bibinfo{author}{\bibfnamefont{P.}~\bibnamefont{Rout}},
  \bibinfo{author}{\bibfnamefont{E.}~\bibnamefont{Maniv}}, \bibnamefont{and}
  \bibinfo{author}{\bibfnamefont{Y.}~\bibnamefont{Dagan}},
  \bibinfo{journal}{Physical Review Letters} \textbf{\bibinfo{volume}{119}},
  \bibinfo{pages}{237002} (\bibinfo{year}{2017}{\natexlab{b}}).

\bibitem[{\citenamefont{Caviglia et~al.}(2008)\citenamefont{Caviglia, Gariglio,
  Reyren, Jaccard, Schneider, Gabay, Thiel, Hammerl, Mannhart, and
  Triscone}}]{caviglia2008electric}
\bibinfo{author}{\bibfnamefont{A.}~\bibnamefont{Caviglia}},
  \bibinfo{author}{\bibfnamefont{S.}~\bibnamefont{Gariglio}},
  \bibinfo{author}{\bibfnamefont{N.}~\bibnamefont{Reyren}},
  \bibinfo{author}{\bibfnamefont{D.}~\bibnamefont{Jaccard}},
  \bibinfo{author}{\bibfnamefont{T.}~\bibnamefont{Schneider}},
  \bibinfo{author}{\bibfnamefont{M.}~\bibnamefont{Gabay}},
  \bibinfo{author}{\bibfnamefont{S.}~\bibnamefont{Thiel}},
  \bibinfo{author}{\bibfnamefont{G.}~\bibnamefont{Hammerl}},
  \bibinfo{author}{\bibfnamefont{J.}~\bibnamefont{Mannhart}}, \bibnamefont{and}
  \bibinfo{author}{\bibfnamefont{J.-M.} \bibnamefont{Triscone}},
  \bibinfo{journal}{Nature} \textbf{\bibinfo{volume}{456}},
  \bibinfo{pages}{624} (\bibinfo{year}{2008}).

\bibitem[{\citenamefont{Kalisky et~al.}(2013)\citenamefont{Kalisky, Spanton,
  Noad, Kirtley, Nowack, Bell, Sato, Hosoda, Xie, Hikita
  et~al.}}]{kalisky2013locally}
\bibinfo{author}{\bibfnamefont{B.}~\bibnamefont{Kalisky}},
  \bibinfo{author}{\bibfnamefont{E.~M.} \bibnamefont{Spanton}},
  \bibinfo{author}{\bibfnamefont{H.}~\bibnamefont{Noad}},
  \bibinfo{author}{\bibfnamefont{J.~R.} \bibnamefont{Kirtley}},
  \bibinfo{author}{\bibfnamefont{K.~C.} \bibnamefont{Nowack}},
  \bibinfo{author}{\bibfnamefont{C.}~\bibnamefont{Bell}},
  \bibinfo{author}{\bibfnamefont{H.~K.} \bibnamefont{Sato}},
  \bibinfo{author}{\bibfnamefont{M.}~\bibnamefont{Hosoda}},
  \bibinfo{author}{\bibfnamefont{Y.}~\bibnamefont{Xie}},
  \bibinfo{author}{\bibfnamefont{Y.}~\bibnamefont{Hikita}},
  \bibnamefont{et~al.}, \bibinfo{journal}{Nat. Mater.}
  \textbf{\bibinfo{volume}{12}}, \bibinfo{pages}{1091} (\bibinfo{year}{2013}).

\bibitem[{\citenamefont{Honig et~al.}(2013)\citenamefont{Honig, Sulpizio,
  Drori, Joshua, Zeldov, and Ilani}}]{IlaniSTOLAODomains}
\bibinfo{author}{\bibfnamefont{M.}~\bibnamefont{Honig}},
  \bibinfo{author}{\bibfnamefont{J.~A.} \bibnamefont{Sulpizio}},
  \bibinfo{author}{\bibfnamefont{J.}~\bibnamefont{Drori}},
  \bibinfo{author}{\bibfnamefont{A.}~\bibnamefont{Joshua}},
  \bibinfo{author}{\bibfnamefont{E.}~\bibnamefont{Zeldov}}, \bibnamefont{and}
  \bibinfo{author}{\bibfnamefont{S.}~\bibnamefont{Ilani}},
  \bibinfo{journal}{Nature Materials} \textbf{\bibinfo{volume}{12}},
  \bibinfo{pages}{1112 EP } (\bibinfo{year}{2013}), \bibinfo{note}{article},
  \urlprefix\url{http://dx.doi.org/10.1038/nmat3810}.

\bibitem[{\citenamefont{Biscaras et~al.}(2014)\citenamefont{Biscaras, Hurand,
  Feuillet-Palma, Rastogi, Budhani, Reyren, Lesne, Lesueur, and
  Bergeal}}]{Biscaras2014}
\bibinfo{author}{\bibfnamefont{J.}~\bibnamefont{Biscaras}},
  \bibinfo{author}{\bibfnamefont{S.}~\bibnamefont{Hurand}},
  \bibinfo{author}{\bibfnamefont{C.}~\bibnamefont{Feuillet-Palma}},
  \bibinfo{author}{\bibfnamefont{A.}~\bibnamefont{Rastogi}},
  \bibinfo{author}{\bibfnamefont{R.~C.} \bibnamefont{Budhani}},
  \bibinfo{author}{\bibfnamefont{N.}~\bibnamefont{Reyren}},
  \bibinfo{author}{\bibfnamefont{E.}~\bibnamefont{Lesne}},
  \bibinfo{author}{\bibfnamefont{J.}~\bibnamefont{Lesueur}}, \bibnamefont{and}
  \bibinfo{author}{\bibfnamefont{N.}~\bibnamefont{Bergeal}},
  \bibinfo{journal}{Scientific Reports} \textbf{\bibinfo{volume}{4}},
  \bibinfo{pages}{6788 EP } (\bibinfo{year}{2014}), \bibinfo{note}{article},
  \urlprefix\url{http://dx.doi.org/10.1038/srep06788}.

\bibitem[{\citenamefont{Davis et~al.}(2017{\natexlab{b}})\citenamefont{Davis,
  Chandrasekhar, Huang, Han, Ariando, and Venkatesan}}]{davis2017anisotropic}
\bibinfo{author}{\bibfnamefont{S.}~\bibnamefont{Davis}},
  \bibinfo{author}{\bibfnamefont{V.}~\bibnamefont{Chandrasekhar}},
  \bibinfo{author}{\bibfnamefont{Z.}~\bibnamefont{Huang}},
  \bibinfo{author}{\bibfnamefont{K.}~\bibnamefont{Han}},
  \bibinfo{author}{\bibnamefont{Ariando}}, \bibnamefont{and}
  \bibinfo{author}{\bibfnamefont{T.}~\bibnamefont{Venkatesan}},
  \bibinfo{journal}{Phys. Rev. B} \textbf{\bibinfo{volume}{95}},
  \bibinfo{pages}{035127} (\bibinfo{year}{2017}{\natexlab{b}}).

\bibitem[{\citenamefont{Davis et~al.}(2017{\natexlab{c}})\citenamefont{Davis,
  Chandrasekhar, Huang, Han, Ariando, and Venkatesan}}]{STOLAO111Holes}
\bibinfo{author}{\bibfnamefont{S.}~\bibnamefont{Davis}},
  \bibinfo{author}{\bibfnamefont{V.}~\bibnamefont{Chandrasekhar}},
  \bibinfo{author}{\bibfnamefont{Z.}~\bibnamefont{Huang}},
  \bibinfo{author}{\bibfnamefont{K.}~\bibnamefont{Han}},
  \bibinfo{author}{\bibnamefont{Ariando}}, \bibnamefont{and}
  \bibinfo{author}{\bibfnamefont{T.}~\bibnamefont{Venkatesan}},
  \bibinfo{journal}{Phys. Rev. B} \textbf{\bibinfo{volume}{95}},
  \bibinfo{pages}{035127} (\bibinfo{year}{2017}{\natexlab{c}}),
  \urlprefix\url{https://link.aps.org/doi/10.1103/PhysRevB.95.035127}.

\bibitem[{\citenamefont{Maniv et~al.}(2015)\citenamefont{Maniv, Ben~Shalom,
  Ron, Mograbi, Palevski, Goldstein, and Dagan}}]{maniv2015strong}
\bibinfo{author}{\bibfnamefont{E.}~\bibnamefont{Maniv}},
  \bibinfo{author}{\bibfnamefont{M.}~\bibnamefont{Ben~Shalom}},
  \bibinfo{author}{\bibfnamefont{A.}~\bibnamefont{Ron}},
  \bibinfo{author}{\bibfnamefont{M.}~\bibnamefont{Mograbi}},
  \bibinfo{author}{\bibfnamefont{A.}~\bibnamefont{Palevski}},
  \bibinfo{author}{\bibfnamefont{M.}~\bibnamefont{Goldstein}},
  \bibnamefont{and} \bibinfo{author}{\bibfnamefont{Y.}~\bibnamefont{Dagan}},
  \bibinfo{journal}{Nat. Commun.} \textbf{\bibinfo{volume}{6}},
  \bibinfo{pages}{8239} (\bibinfo{year}{2015}).

\bibitem[{\citenamefont{Bell et~al.}(2009)\citenamefont{Bell, Harashima,
  Kozuka, Kim, Kim, Hikita, and Hwang}}]{bellDominant}
\bibinfo{author}{\bibfnamefont{C.}~\bibnamefont{Bell}},
  \bibinfo{author}{\bibfnamefont{S.}~\bibnamefont{Harashima}},
  \bibinfo{author}{\bibfnamefont{Y.}~\bibnamefont{Kozuka}},
  \bibinfo{author}{\bibfnamefont{M.}~\bibnamefont{Kim}},
  \bibinfo{author}{\bibfnamefont{B.~G.} \bibnamefont{Kim}},
  \bibinfo{author}{\bibfnamefont{Y.}~\bibnamefont{Hikita}}, \bibnamefont{and}
  \bibinfo{author}{\bibfnamefont{H.~Y.} \bibnamefont{Hwang}},
  \bibinfo{journal}{Phys. Rev. Lett.} \textbf{\bibinfo{volume}{103}},
  \bibinfo{pages}{226802} (\bibinfo{year}{2009}),
  \urlprefix\url{https://link.aps.org/doi/10.1103/PhysRevLett.103.226802}.

\bibitem[{\citenamefont{Kapitulnik et~al.}(2017)\citenamefont{Kapitulnik,
  Kivelson, and Spivak}}]{kapitulnik2017anomalousReview}
\bibinfo{author}{\bibfnamefont{A.}~\bibnamefont{Kapitulnik}},
  \bibinfo{author}{\bibfnamefont{S.~A.} \bibnamefont{Kivelson}},
  \bibnamefont{and} \bibinfo{author}{\bibfnamefont{B.}~\bibnamefont{Spivak}},
  \bibinfo{journal}{arXiv preprint arXiv:1712.07215}  (\bibinfo{year}{2017}).

\bibitem[{\citenamefont{Sac{\'e}p{\'e}
  et~al.}(2011)\citenamefont{Sac{\'e}p{\'e}, Dubouchet, Chapelier, Sanquer,
  Ovadia, Shahar, Feigel'man, and Ioffe}}]{ShaharIslandsSTM}
\bibinfo{author}{\bibfnamefont{B.}~\bibnamefont{Sac{\'e}p{\'e}}},
  \bibinfo{author}{\bibfnamefont{T.}~\bibnamefont{Dubouchet}},
  \bibinfo{author}{\bibfnamefont{C.}~\bibnamefont{Chapelier}},
  \bibinfo{author}{\bibfnamefont{M.}~\bibnamefont{Sanquer}},
  \bibinfo{author}{\bibfnamefont{M.}~\bibnamefont{Ovadia}},
  \bibinfo{author}{\bibfnamefont{D.}~\bibnamefont{Shahar}},
  \bibinfo{author}{\bibfnamefont{M.}~\bibnamefont{Feigel'man}},
  \bibnamefont{and} \bibinfo{author}{\bibfnamefont{L.}~\bibnamefont{Ioffe}},
  \bibinfo{journal}{Nature Physics} \textbf{\bibinfo{volume}{7}},
  \bibinfo{pages}{239 EP } (\bibinfo{year}{2011}), \bibinfo{note}{article},
  \urlprefix\url{http://dx.doi.org/10.1038/nphys1892}.

\bibitem[{\citenamefont{Steiner and Kapitulnik}(2005)}]{KapitulnikPhysicaC}
\bibinfo{author}{\bibfnamefont{M.}~\bibnamefont{Steiner}} \bibnamefont{and}
  \bibinfo{author}{\bibfnamefont{A.}~\bibnamefont{Kapitulnik}},
  \bibinfo{journal}{Physica C: Superconductivity}
  \textbf{\bibinfo{volume}{422}}, \bibinfo{pages}{16 } (\bibinfo{year}{2005}),
  ISSN \bibinfo{issn}{0921-4534},
  \urlprefix\url{http://www.sciencedirect.com/science/article/pii/S0921453405000894}.

\bibitem[{\citenamefont{Ron et~al.}(2014)\citenamefont{Ron, Maniv, Graf, Park,
  and Dagan}}]{ron2014anomalous}
\bibinfo{author}{\bibfnamefont{A.}~\bibnamefont{Ron}},
  \bibinfo{author}{\bibfnamefont{E.}~\bibnamefont{Maniv}},
  \bibinfo{author}{\bibfnamefont{D.}~\bibnamefont{Graf}},
  \bibinfo{author}{\bibfnamefont{J.-H.} \bibnamefont{Park}}, \bibnamefont{and}
  \bibinfo{author}{\bibfnamefont{Y.}~\bibnamefont{Dagan}},
  \bibinfo{journal}{Phys. Rev. Lett.} \textbf{\bibinfo{volume}{113}},
  \bibinfo{pages}{216801} (\bibinfo{year}{2014}).

\bibitem[{\citenamefont{Marziali~Berm\'udez
  et~al.}(2017)\citenamefont{Marziali~Berm\'udez, Louden, Eskildsen, Dewhurst,
  Bekeris, and Pasquini}}]{VortexPinningHysteresis}
\bibinfo{author}{\bibfnamefont{M.}~\bibnamefont{Marziali~Berm\'udez}},
  \bibinfo{author}{\bibfnamefont{E.~R.} \bibnamefont{Louden}},
  \bibinfo{author}{\bibfnamefont{M.~R.} \bibnamefont{Eskildsen}},
  \bibinfo{author}{\bibfnamefont{C.~D.} \bibnamefont{Dewhurst}},
  \bibinfo{author}{\bibfnamefont{V.}~\bibnamefont{Bekeris}}, \bibnamefont{and}
  \bibinfo{author}{\bibfnamefont{G.}~\bibnamefont{Pasquini}},
  \bibinfo{journal}{Phys. Rev. B} \textbf{\bibinfo{volume}{95}},
  \bibinfo{pages}{104505} (\bibinfo{year}{2017}),
  \urlprefix\url{https://link.aps.org/doi/10.1103/PhysRevB.95.104505}.

\bibitem[{\citenamefont{Biscaras et~al.}(2013)\citenamefont{Biscaras, Bergeal,
  Hurand, Feuillet-Palma, Rastogi, Budhani, Grilli, Caprara, and
  Lesueur}}]{Lesueur2013multiplecriticality}
\bibinfo{author}{\bibfnamefont{J.}~\bibnamefont{Biscaras}},
  \bibinfo{author}{\bibfnamefont{N.}~\bibnamefont{Bergeal}},
  \bibinfo{author}{\bibfnamefont{S.}~\bibnamefont{Hurand}},
  \bibinfo{author}{\bibfnamefont{C.}~\bibnamefont{Feuillet-Palma}},
  \bibinfo{author}{\bibfnamefont{A.}~\bibnamefont{Rastogi}},
  \bibinfo{author}{\bibfnamefont{R.~C.} \bibnamefont{Budhani}},
  \bibinfo{author}{\bibfnamefont{M.}~\bibnamefont{Grilli}},
  \bibinfo{author}{\bibfnamefont{S.}~\bibnamefont{Caprara}}, \bibnamefont{and}
  \bibinfo{author}{\bibfnamefont{J.}~\bibnamefont{Lesueur}},
  \bibinfo{journal}{Nature Materials} \textbf{\bibinfo{volume}{12}},
  \bibinfo{pages}{542 EP } (\bibinfo{year}{2013}), \bibinfo{note}{article},
  \urlprefix\url{http://dx.doi.org/10.1038/nmat3624}.

\bibitem[{\citenamefont{Mehta et~al.}(2014)\citenamefont{Mehta, Dikin, Bark,
  Ryu, Folkman, Eom, and Chandrasekhar}}]{Mehta2014Magnetic}
\bibinfo{author}{\bibfnamefont{M.~M.} \bibnamefont{Mehta}},
  \bibinfo{author}{\bibfnamefont{D.~A.} \bibnamefont{Dikin}},
  \bibinfo{author}{\bibfnamefont{C.~W.} \bibnamefont{Bark}},
  \bibinfo{author}{\bibfnamefont{S.}~\bibnamefont{Ryu}},
  \bibinfo{author}{\bibfnamefont{C.~M.} \bibnamefont{Folkman}},
  \bibinfo{author}{\bibfnamefont{C.~B.} \bibnamefont{Eom}}, \bibnamefont{and}
  \bibinfo{author}{\bibfnamefont{V.}~\bibnamefont{Chandrasekhar}},
  \bibinfo{journal}{Phys. Rev. B} \textbf{\bibinfo{volume}{90}},
  \bibinfo{pages}{100506} (\bibinfo{year}{2014}),
  \urlprefix\url{https://link.aps.org/doi/10.1103/PhysRevB.90.100506}.

\bibitem[{\citenamefont{Shen et~al.}(2016)\citenamefont{Shen, Xing, Wang, Liu,
  Fu, Zhang, He, Xie, Lin, Nie et~al.}}]{GriffithsSTOLAO}
\bibinfo{author}{\bibfnamefont{S.}~\bibnamefont{Shen}},
  \bibinfo{author}{\bibfnamefont{Y.}~\bibnamefont{Xing}},
  \bibinfo{author}{\bibfnamefont{P.}~\bibnamefont{Wang}},
  \bibinfo{author}{\bibfnamefont{H.}~\bibnamefont{Liu}},
  \bibinfo{author}{\bibfnamefont{H.}~\bibnamefont{Fu}},
  \bibinfo{author}{\bibfnamefont{Y.}~\bibnamefont{Zhang}},
  \bibinfo{author}{\bibfnamefont{L.}~\bibnamefont{He}},
  \bibinfo{author}{\bibfnamefont{X.~C.} \bibnamefont{Xie}},
  \bibinfo{author}{\bibfnamefont{X.}~\bibnamefont{Lin}},
  \bibinfo{author}{\bibfnamefont{J.}~\bibnamefont{Nie}}, \bibnamefont{et~al.},
  \bibinfo{journal}{Phys. Rev. B} \textbf{\bibinfo{volume}{94}},
  \bibinfo{pages}{144517} (\bibinfo{year}{2016}),
  \urlprefix\url{https://link.aps.org/doi/10.1103/PhysRevB.94.144517}.

\bibitem[{\citenamefont{Sherman et~al.}(2012)\citenamefont{Sherman, Kopnov,
  Shahar, and Frydman}}]{ShaharIslandsJunction}
\bibinfo{author}{\bibfnamefont{D.}~\bibnamefont{Sherman}},
  \bibinfo{author}{\bibfnamefont{G.}~\bibnamefont{Kopnov}},
  \bibinfo{author}{\bibfnamefont{D.}~\bibnamefont{Shahar}}, \bibnamefont{and}
  \bibinfo{author}{\bibfnamefont{A.}~\bibnamefont{Frydman}},
  \bibinfo{journal}{Phys. Rev. Lett.} \textbf{\bibinfo{volume}{108}},
  \bibinfo{pages}{177006} (\bibinfo{year}{2012}),
  \urlprefix\url{https://link.aps.org/doi/10.1103/PhysRevLett.108.177006}.

\bibitem[{\citenamefont{Emery and Kivelson}(1995)}]{KivelsonEmery1995}
\bibinfo{author}{\bibfnamefont{V.~J.} \bibnamefont{Emery}} \bibnamefont{and}
  \bibinfo{author}{\bibfnamefont{S.~A.} \bibnamefont{Kivelson}},
  \bibinfo{journal}{Nature} \textbf{\bibinfo{volume}{374}}, \bibinfo{pages}{434
  EP } (\bibinfo{year}{1995}),
  \urlprefix\url{http://dx.doi.org/10.1038/374434a0}.

\bibitem[{\citenamefont{Cheng et~al.}(2015)\citenamefont{Cheng, Tomczyk, Lu,
  Veazey, Huang, Irvin, Ryu, Lee, Eom, Hellberg et~al.}}]{Levy2015QD}
\bibinfo{author}{\bibfnamefont{G.}~\bibnamefont{Cheng}},
  \bibinfo{author}{\bibfnamefont{M.}~\bibnamefont{Tomczyk}},
  \bibinfo{author}{\bibfnamefont{S.}~\bibnamefont{Lu}},
  \bibinfo{author}{\bibfnamefont{J.~P.} \bibnamefont{Veazey}},
  \bibinfo{author}{\bibfnamefont{M.}~\bibnamefont{Huang}},
  \bibinfo{author}{\bibfnamefont{P.}~\bibnamefont{Irvin}},
  \bibinfo{author}{\bibfnamefont{S.}~\bibnamefont{Ryu}},
  \bibinfo{author}{\bibfnamefont{H.}~\bibnamefont{Lee}},
  \bibinfo{author}{\bibfnamefont{C.-B.} \bibnamefont{Eom}},
  \bibinfo{author}{\bibfnamefont{C.~S.} \bibnamefont{Hellberg}},
  \bibnamefont{et~al.}, \bibinfo{journal}{Nature}
  \textbf{\bibinfo{volume}{521}}, \bibinfo{pages}{196} (\bibinfo{year}{2015}).

\bibitem[{\citenamefont{Richter et~al.}(2013)\citenamefont{Richter, Boschker,
  Dietsche, Fillis-Tsirakis, Jany, Loder, Kourkoutis, Muller, Kirtley,
  Schneider et~al.}}]{Manhart2013Tunneling}
\bibinfo{author}{\bibfnamefont{C.}~\bibnamefont{Richter}},
  \bibinfo{author}{\bibfnamefont{H.}~\bibnamefont{Boschker}},
  \bibinfo{author}{\bibfnamefont{W.}~\bibnamefont{Dietsche}},
  \bibinfo{author}{\bibfnamefont{E.}~\bibnamefont{Fillis-Tsirakis}},
  \bibinfo{author}{\bibfnamefont{R.}~\bibnamefont{Jany}},
  \bibinfo{author}{\bibfnamefont{F.}~\bibnamefont{Loder}},
  \bibinfo{author}{\bibfnamefont{L.~F.} \bibnamefont{Kourkoutis}},
  \bibinfo{author}{\bibfnamefont{D.~A.} \bibnamefont{Muller}},
  \bibinfo{author}{\bibfnamefont{J.~R.} \bibnamefont{Kirtley}},
  \bibinfo{author}{\bibfnamefont{C.~W.} \bibnamefont{Schneider}},
  \bibnamefont{et~al.}, \bibinfo{journal}{Nature}
  \textbf{\bibinfo{volume}{502}}, \bibinfo{pages}{528 EP }
  (\bibinfo{year}{2013}),
  \urlprefix\url{http://dx.doi.org/10.1038/nature12494}.

\end{thebibliography}

\bigskip
\noindent\textbf{Acknowledgements}

We are indebted to Aharon Kapitulnik, Dan Shahar, Moshe Goldstein, Alexander Palevski, Efrat Shimshoni, Tsofar Maniv, Igor Bormistrov, 	
Pratap Raychaudhuri, Amnon Aharoni, Ora Entin-Wohlman, Tal Levinson, Adam Doron and Idan Tamir for useful discussions. This work has been supported by the Israel Science Foundation under grant 382/17 and the Bi-national science foundation under grant 2014047. A portion of this work was performed at the National High Magnetic Field Laboratory, which is supported by National Science Foundation Cooperative Agreement No. DMR-1157490 and the State of Florida.

\bigskip
\noindent\textbf{Author contributions}

M.M., P.K.R. and E.M. contributed equally to this work. M.M., P.K.R., E.M. and Y.D. performed the transport measurements and analyzed the data. D.G. and J.P. performed the measurements at the NHMFL. P.K.R and E.M. prepared the samples. All authors discussed the data and wrote the paper.

\end{document}